\documentclass{ws-ijmpcs_gtb}

\begin{document}

\markboth{Geoffrey T.~Bodwin}
{NRQCD Factorization and Quarkonium Production at Hadron-Hadron
and $ep$ Colliders}

%
\catchline{}{}{}{}{}
%

\title{
NRQCD FACTORIZATION AND QUARKONIUM PRODUCTION AT HADRON-HADRON
AND $ep$ COLLIDERS}

\author{GEOFFREY T.~BODWIN}

\address{High Energy Physics Division, Argonne National Laboratory, 9700 
South Cass Avenue,\\
Argonne, Illinois 60439, USA\\
gtb@hep.anl.gov}

\maketitle

\begin{history}
\hbox{}\hfill\break
\end{history}

\begin{abstract}
I review the NRQCD factorization approach for quarkonium production and
compare its predictions with measurements in hadron-hadron
and $ep$ collisions.
\end{abstract}

\ccode{PACS numbers: 12.38.-t,14.40.Gx,12.39.St}
\hfill\break
I refer the reader to Ref.~\refcite{Brambilla:2010cs} as a supplement to
this brief account.

\section{FACTORIZATION OF THE INCLUSIVE QUARKONIUM PRODUCTION CROSS SECTION}

In heavy-quarkonium production in hard-scattering processes, two large
momentum scales appear: the heavy-quark mass $m$ and the typical
momentum transfer in the hard scattering, which I will denote
generically by $p_T$. One would like to separate the perturbative physics
at these large momentum scales from the physics at smaller momentum
scales that is associated with nonperturbative heavy-quarkonium
bound-state dynamics. It has been conjectured\cite{Bodwin:1994jh} that,
for the inclusive quarkonium production cross section at $p_T\gg m$, one
can achieve such a separation and that one can write the cross section
in the following factorized form:
\begin{equation}
\sigma(H)=\sum_n F_n\langle 0|
{\cal O}_n^H|0\rangle.
\label{factorization}
\end{equation}
The $F_n$ are ``short-distance coefficients.'' They are essentially the
process-dependent partonic hard-scattering cross sections convolved with
the parton distributions. The partonic hard-scattering cross sections
depend only on the large scales $m$ and $p_T$, and they have an
expansion in powers of $\alpha_s$. The quantities $\langle 0| {\cal
O}_n^H(\Lambda)|0\rangle$ are long-distance matrix elements (LDMEs) that
are formulated in terms of the effective field theory nonrelativistic
QCD (NRQCD). They give the probability for a heavy $Q\bar Q$ pair with a
certain set of quantum numbers to evolve into a heavy quarkonium $H$.
The LDMEs have a known scaling with $v$, the heavy-quark velocity in the
quarkonium rest frame. Hence, the factorization formula in
Eq.~(\ref{factorization}) is a double expansion in powers of $\alpha_s$
and $v$. In practice, the sum in Eq.~(\ref{factorization}) is truncated
at a finite order in $v$. The LDMEs are believed to be universal
(process independent)---an assumption that gives the factorization
formula much of its predictive power.

A key feature of NRQCD factorization is that quarkonium production can
occur through color-octet, as well as color-singlet, $Q\bar Q$ states.
If one drops all of the color-octet contributions and retains only
the leading color-singlet contribution, then one obtains the
color-singlet model (CSM). The CSM is inconsistent for production
processes beyond leading order in $v$, for example, production through
$P$-wave $Q\bar Q$ channels, because it leads to uncanceled infrared
divergences.

\subsection{Status of a Proof of Factorization}

A proof of the NRQCD factorization formula in Eq.~(\ref{factorization})
is complicated because gluons can dress the basic production process in
ways that apparently violate factorization, for example, by connecting
one incoming hadron to another or by connecting an incoming hadron to
the quarkonium. A proof of factorization would involve a demonstration
that diagrams in each order in $\alpha_s$ can be re-organized so that
(1) all soft singularities cancel or can be absorbed into NRQCD matrix
elements and (2) all collinear singularities and spectator interactions
can be absorbed into parton distributions.

Nayak, Qiu, and Sterman pointed out that, in order to make the
color-octet NRQCD LDMEs gauge invariant, one must be modify them from
the original definition\cite{Bodwin:1994jh} by including eikonal lines
(path-ordered exponentials of path integrals of the gauge field) that
run from the points at which the $Q\bar Q$ pair is created to
infinity.\cite{Nayak:2005rw,Nayak:2005rt} These eikonal lines are
essential at two-loop order in order to allow certain soft contributions
to be absorbed into the LDMEs. However, they do not affect existing
phenomenology, which is at tree order or one-loop order in the case of
the color-octet channels. A key difficulty in proving factorization to
all orders in $\alpha_s$ is the treatment of gluons that have momenta of
order $m$ in the quarkonium rest frame. Such gluons can communicate with
the $Q\bar Q$ pair through the exchange of soft gluons, which lead to
infrared divergences. One could absorb these infrared divergences into
the LDMEs by treating the gluon with momenta of order $m$ as an eikonal
line in each LDME. However, such a treatment would preserve the
universality of the LDMEs only if they are independent of the direction
of the eikonal line (the direction of the gluon). In an explicit
calculation at two-loop order, the dependence on the direction of the
eikonal line cancels.\cite{Nayak:2005rw,Nayak:2005rt} However, it is not
known if this cancellation generalizes to all orders. An all-orders
proof of factorization is essential because non-factorizing soft-gluon
contributions, for which $\alpha_s$ is not small, could be important
numerically at any order.

Nayak, Qiu, Sterman have also pointed out that, if an additional heavy
quark is approximately co-moving with the $Q\bar Q$ pair that evolves
into the quarkonium, then there can be soft color exchanges between the
heavy quark and the $Q\bar Q$ pair.\cite{Nayak:2007mb} This process does
not fit into the NRQCD factorization picture because it involves LDMEs
that contain additional heavy quarks beyond the $Q\bar Q$ pair that
evolves into the quarkonium. The process is nonperturbative and,
therefore, its rate cannot be calculated reliably. However, the process
can be identified experimentally. Its signature is an excess of
heavy-flavor mesons in a cone around the quarkonium of angular size of
order $mv$ divided by the quarkonium momentum.

\section{Quarkonium Production in Hadron-Hadron Collisions}

It has been known for many years that the cross sections differential in
$p_T$ for the production of the $J/\psi$, $\psi(2S)$, $\chi_{cJ}$, and
$\Upsilon$ states at the Tevatron can be fit well by the predictions of
NRQCD factorization at leading order (LO) in $\alpha_s$. The existing
fits have included the contributions from the LDMEs of leading order in
$v$ in the color-singlet and color-octet channels, and, in the $S$-wave
case, the two independent color-octet LDMEs of the first subleading
order in $v$. The color-singlet LDMEs can be fixed by decay rates, but
the color-octet LDMEs are treated as free parameters. While NRQCD
factorization passes the minimal test of fitting the Tevatron cross
sections, it is important to make more stringent tests by using the
fitted values of the LDMEs to predict quarkonium production rates in
other processes and to predict values for other observables, such as
quarkonium polarizations.

At LO in $\alpha_s$, NRQCD factorization predicts that prompt $J/\psi$
and $\psi(2S)$ polarizations at the Tevatron should be substantially
transverse for values of the quarkonium $p_T$ that are greater than
$m_c$.\cite{Braaten:1999qk} This prediction is compatible, within
uncertainties, with the CDF Run~I measurement,\cite{Affolder:2000nn} but
is incompatible with the CDF Run~II measurement.\cite{Abulencia:2007us}
The Run~I and Run~II results are also mutually inconsistent, but the CDF
collaboration endorses the Run~II results. The CDF Run~II data for
the $\psi(2S)$ polarization\cite{Abulencia:2007us} are also incompatible
with the LO NRQCD prediction. In the case of $\Upsilon$ polarization, the
CDF\cite{CDF-pol-note} and D0\cite{Abazov:2008za} measurements are
incompatible with each other and with the LO NRQCD
prediction.\cite{Braaten:2000gw} The discrepancies between theory and
experiment with regard to polarization present a significant challenge
to our understanding of the mechanisms of quarkonium production.
Comparisons of predictions for the polarization of the $J/\psi$ with
experimental measurements are greatly complicated by the presence of
feeddown contributions from the $\psi(2S)$ and $\chi_{cJ}$ states.
Measurements the polarization of the $J/\psi$ in direct production would
be of considerable help understanding the production mechanisms.

In the case of quarkonium production in the color-singlet channel,
complete corrections at next-to-leading order (NLO) in $\alpha_s$ and
real-gluon corrections at next-to-next-to-leading order in $\alpha_s$
(NNLO$^*$) have been calculated. These calculations have revealed the
surprising result that, at large $p_T$, higher-order contributions can
be enhanced by more than an order of magnitude relative to lower-order
contributions because the higher-order contributions fall off less
rapidly as $p_T$ increases.\cite{Campbell:2007ws,Artoisenet:2008fc} In
the case of prompt $J/\psi$ production at CDF\cite{Acosta:2004yw}, the
NNLO$^*$ color-singlet contribution\cite{Artoisenet:2008zzc} lies well
below the data, suggesting that most of the cross section comes from a
color-octet contribution. In the case of prompt $\Upsilon$ production at
CDF,\cite{Acosta:2001gv} the NNLO$^*$ color-singlet
contribution\cite{Artoisenet:2008fc} could explain the data by itself.
However, the large theoretical uncertainties still allow the presence of
a substantial color-octet contribution. In the case of $J/\psi$
production at STAR\cite{Abelev:2009qaa} and at
PHENIX,\cite{Adare:2006kf} the NLO and NNLO$^*$ color-singlet corrections
are large,\cite{Lansberg:2010vq} and the upper limit of the large
uncertainty band for the NNLO$^*$ prediction is close to the STAR data
at large $p_T$.

NLO calculations of the ${}^1S_0$ and ${}^3S_1$ color-octet
contributions to $S$-wave quarkonium production at the Tevatron and the
LHC show that the corrections to $J/\psi$ and $\psi(2S)$ production are
less than $25\%$,\cite{Gong:2008ft} while the corrections to $\Upsilon$
production are less than $40\%$.\cite{Gong:2010bk} Recently, two groups
have completed the first NLO calculations that include all of the
color-octet channels through the first subleading order in
$v$.\cite{Ma:2010yw,Butenschoen:2010rq} The numerical results of the two
groups for the short-distance coefficients are in agreement and show a
very large negative $K$ factor in the ${}^3P_J$ channels. In
Ref.~\refcite{Butenschoen:2010rq}, the LDMEs were extracted in a fit to
both the CDF Run~II data\cite{Acosta:2004yw} and the H1 photoproduction
data,\cite{Adloff:2002ex,Aaron:2010gz} and the resulting LDMEs are not
qualitatively different in size from those that were obtained in LO
fits to the Tevatron data. In Ref.~\refcite{Ma:2010yw}, fits were made
only to the CDF Run~II data,\cite{Acosta:2004yw} and two
linear combinations of LDMEs were determined. The fitting procedures
that were used in Refs.~\refcite{Ma:2010yw} and
\refcite{Butenschoen:2010rq} differ in a number of respects, beyond the
inclusion/exclusion of the H1 data, and the LDMEs that were obtained
differ substantially, in one case by an order of magnitude. Clearly, it
will be necessary to examine the validity of the assumptions that are
implicit in the fitting procedures before any conclusions can be drawn
about the sizes of the LDMEs. In Refs.~\refcite{Ma:2010yw} and
\refcite{Butenschoen:2010rq}, the extracted values of the LDMEs were
used to make predictions for the prompt $J/\psi$ cross section
differential in $p_T$ at CMS,\cite{CMS} and good agreement was obtained.
In Ref.~\refcite{Butenschoen:2010rq}, a prediction was also made for the
$J/\psi$ cross section at PHENIX,\cite{Adare:2009js} and good agreement
was again obtained. The fact that good agreement with the CDF and CMS
data was obtained using very different values of the LDMEs suggests that
the hadroproduction cross section differential in $p_T$ is rather
insensitive to the details of the production mechanism.

NLO and NNLO$^*$ calculations of the color-singlet contribution to
the $J/\psi$ and $\Upsilon$ polarizations reveal that they change from
transverse to longitudinal once one goes beyond
LO.\cite{Artoisenet:2008fc,Gong:2008sn,Gong:2008hk} NLO corrections to the
color-octet $S$-wave contributions to the $J/\psi$ polarization produce
only small effects.\cite{Gong:2008ft} Given the large sizes of the NLO
corrections to the rates from color-octet ${}^3P_J$ channels, it is
clearly necessary to compute the contributions to the polarization from
these channels before a definitive comparison can be made between NRQCD
factorization predictions and experiment.

The large corrections that occur at NLO and NNLO$^*$ cast some doubt on
the convergence of the perturbation series. As I have mentioned, the
large corrections arise, at least in part, because the NLO and NNLO$^*$
contributions in certain $Q\bar Q$ channels fall more slowly with $p_T$
than do the lower-order contributions. The $p_T$ distribution can fall
no more slowly than $1/p_T^4$. This dependence is achieved in the
color-singlet channel at NNLO$^*$ and in the color-octet channels at
NLO. Hence, no further kinematic enhancements are expected beyond these
orders. However, the theoretical uncertainties are large in these orders
because the kinematically enhanced contributions are, in effect,
computed at the Born level. Kang, Qiu, and Sterman have suggested that
this difficulty might be overcome by using fragmentation-function
methods to reorganize the perturbative calculation according to the
$p_T$ dependence of the contributions.\cite{kang-qiu-sterman} Such a
reorganization might make it possible to compute more accurately 
the contributions that are most important numerically.

\section{Quarkonium production in $ep$ collisions}

Until recently, it had been believed that NLO color-singlet
contributions account quite well for the $J/\psi$ photoproduction cross
sections that have been measured at HERA, leaving little room for a
color-octet contribution.\cite{Kramer:1994zi,Kramer:1995nb} However, a
new calculation of the NLO color-singlet
contribution,\cite{Artoisenet:2009xh} while confirming the analytic
results of the previous calculations, reveals that a more reasonable
choice of the renormalization/factorization scale yields a much smaller
numerical value for the color-singlet contribution, thereby opening the
possibility that there is an appreciable color-octet contribution. This
possibility was substantiated by a complete NLO analysis, including
color-octet contributions,\cite{Butenschoen:2010rq,Butenschoen:2009zy}
in which it was found that color-octet contributions are necessary in
order to obtain agreement with the H1
data.\cite{Adloff:2002ex,Aaron:2010gz} While the NLO cross section
differential in $p_T$ is in good agreement with the H1 data, the NLO
cross section differential in the $J/\psi$ energy fraction $z$ deviates
from the data at low and high $z$. However, these deviations are
understood as arising from the presence of uncalculated
resolved-photoproduction contributions at small $z$ and the failures of
convergence of the NRQCD velocity expansion and the perturbation
expansion near $z=1$, both of which require a resummation in order to
obtain reliable predictions. 

The color-singlet contribution to the $J/\psi$ polarization in
photoproduction is strongly affected by QCD
corrections,\cite{Artoisenet:2009xh,Chang:2009uj} changing from largely
transverse at LO to largely longitudinal at NLO. Comparisons of the NLO
color-singlet contribution with the H1 and Zeus
data\cite{Jungst:2008ip,Chekanov:2009br} show that the color-singlet
contribution alone cannot explain the observed polarization. A complete
NLO calculation, including color-octet contributions, would be necessary
in order to make a meaningful comparison between the predictions of NRQCD
factorization and the data.

An LO calculation of $J/\psi$ production in deep-inelastic scattering at HERA 
is generally in good agreement with the data.\cite{Kniehl:2001tk} 
However, given the importance of NLO corrections in other quarkonium 
production processes, it would be prudent to wait for the calculation of 
the NLO corrections to deep-inelastic scattering before drawing any 
conclusions.

%

\section*{Acknowledgments}

I thank Bernd Kniehl, Mathias Butensch\"on, Kuang-Ta Chao, and Yan-Qing
Ma for illuminating discussions. This work was supported by the U.S.
Department of Energy, Division of High Energy Physics, under Contract
No.~DE-AC02-06CH11357.

\end{document}